\documentclass[sigconf]{acmart}
\usepackage{graphicx} 
\usepackage{longtable}
\usepackage{array}
\usepackage{comment}
\usepackage[many]{tcolorbox} 
\usepackage{amsmath}
\usepackage{tabularx}  
\usepackage{enumitem} 
\usepackage{multirow}
\usepackage{makecell,booktabs}
\usepackage[switch]{lineno} 
\usepackage{caption} 
\usepackage{subcaption}
\usepackage{balance}

\setlist{nosep,leftmargin=*}
\renewcommand{\arraystretch}{0.9} 
\acmConference[SIGCITE '25]{Proceedings of the 26th ACM Annual Conference on Cybersecurity and Information Technology Education (ACM SIGCITE 2025)}{November 6--8, 2025}{Sacramento, California, USA}
\acmYear{2025}
\copyrightyear{2025}
\acmISBN{978-1-4503-XXXX-X/25/11}  
\acmDOI{xxx} 
\setcopyright{none}

\title{ChatGPT in Introductory Programming: Counterbalanced Evaluation of Code Quality, Conceptual Learning, and Student Perceptions}

\author{Shiza Andleeb}
\affiliation{%
  \institution{The University of Alabama}
  \city{Tuscaloosa}
  \state{Alabama}
  \country{USA}
}
\email{sandleeb@crimson.ua.edu}

\author{Brandon Kantorski}
\affiliation{%
  \institution{The University of Alabama}
  \city{Tuscaloosa}
  \state{Alabama}
  \country{USA}
}
\email{bfkantorski@crimson.ua.edu}

\author{Jeffrey Carver}
\affiliation{%
  \institution{The University of Alabama}
  \city{Tuscaloosa}
  \state{Alabama}
  \country{USA}
}
\email{carver@cs.ua.edu}



\begin{document}

\begin{abstract}
\noindent\textbf{Background:} Large language models (LLMs) such as ChatGPT are increasingly used in introductory programming courses to provide real-time code generation, debugging, and explanations. 
While these tools can boost productivity and code quality, concerns remain about over-reliance and potential impacts on conceptual learning.

\noindent\textbf{Objective:} 
To investigate how ChatGPT access affects code quality, conceptual understanding, task completion times, and student perceptions in a CS1 course.

\noindent\textbf{Methods:} 
We conducted a counterbalanced, quasi-experimental study in which students alternated between ChatGPT and non-ChatGPT conditions across two programming assignments in C (functions and structures). 
We evaluated their code submissions using multidimensional rubrics, conceptual post-surveys, and task completion time. 

\noindent\textbf{Results:} 
Students who had access to ChatGPT produced significantly higher rubric scores for code quality and completed tasks in less time compared to those without access. 
However, gains in conceptual understanding were mixed—lower for the functions topic but higher for the structures topic.
Students reported positive experiences with ChatGPT, citing its value for debugging and practice, while expressing concerns about accuracy and long-term skill development.

\noindent\textbf{Conclusions:} 
ChatGPT can enhance code quality and efficiency for novice programmers, but may not uniformly improve conceptual understanding. 
Structured integration and complementary instructional strategies are recommended to foster independent problem-solving skills.

\end{abstract}

\maketitle 
\renewcommand\linenumberfont{\normalfont\tiny\color{red}}

\section{Introduction}
Artificial Intelligence (AI) is reshaping education, especially in computer science, through tools like intelligent tutors, automated feedback, and personalized learning systems. 
Among the most impactful innovations are large language models (LLMs), such as ChatGPT, which provide real-time support in debugging, explanation, and code generation. 
ChatGPT has rapidly entered computing classrooms, where students use it to clarify programming concepts and improve code quality.
While early evidence suggests LLMs may boost productivity and reduce cognitive load, concerns remain about students' overreliance on them and their impact on student learning. 
Current literature often focuses on student perceptions or use cases for ChatGPT, with limited empirical evidence on how ChatGPT affects learning outcomes, particularly in foundational programming courses.

This work addresses that gap by studying the impacts of ChatGPT on beginning CS students, in terms of student performance on assignments, quality of code developed by students, efficiency of the development process, and student learning of fundamental programming concepts.
We employ a counterbalanced, quasi-experimental design in a CS1 course that utilizes the C programming language.
We employed a counterbalanced, quasi-experiment to examine these effects by comparing student performance with and without access to ChatGPT.
We chose CS1 because it is a gateway course where students often struggle with core programming concepts, making it a critical context to evaluate the potential benefits and risks of ChatGPT.
The overall goal of this study is to answer the following question.

\begin{tcolorbox}[enhanced,drop shadow]
\textbf{\textit{How does access to ChatGPT influence learning and performance in foundational programming courses?}}
\end{tcolorbox}

Our work contributes to the field of computing education research in three ways.
First, we employed a counterbalanced experimental design in which both groups alternated between ChatGPT and non-ChatGPT conditions, reducing order effects that are often overlooked in prior AI-in-education studies. 
Second, a validated, multi-dimensional code-quality rubric for CS1 programming tasks (covering correctness, efficiency, code quality, and warning/error analysis). 
The complete rubric, with scoring anchors and examples, is included in this paper to support replication and adaptation in other instructional and research contexts.
Third, it offers empirical insights into how students engage with generative AI in foundational C programming topics, specifically functions and structures.
These contributions enable us to reveal nuanced effects of ChatGPT, showing that while it accelerates task completion and supports procedural coding, its benefits for conceptual mastery are more limited insights that previous single-measure studies could not capture.

\label{introduction}

\section{Related Work}
Recent studies highlight the growing role of AI in education, particularly through tools such as intelligent tutoring systems, automated graders, and virtual assistants that provide personalized instruction, real-time feedback, and enhanced student engagement~\cite{william2024, ArialHanZhou2024, MousaRaneemEmadSigcse2024, YingXieSIGTE2023}.
Educators regularly use code quality, which includes correctness, readability, efficiency, and error handling, to assess student programming skills~\cite{yuankaiICSE_SEET2024, IshikaJoshiSIGCSE2024}.
ChatGPT can help students reduce syntactic errors and improve code readability, but its impact on deeper problem-solving and structural quality is inconsistent without scaffolding~\cite{stephanieyangICER2024, RongxinCarterSIGCSE2024, oliveira2023chatgpt, ghimire2024coding}.
Popovici~\cite{popovici2023chatgpt} demonstrated ChatGPT’s value in code review rather than in code generation, reinforcing its potential for providing formative feedback.
Despite this result, few studies have evaluated AI’s effect on code quality using multidimensional rubrics across multiple topics, such as functions and structures, in a controlled design.
This gap leads to our first research question.

\begin{tcolorbox}[enhanced,drop shadow]
\textbf{RQ1: Does using ChatGPT improve the quality of students’ code in lab tasks as measured by our evaluation rubrics?}
\end{tcolorbox}

One of the frequently cited benefits of AI is reducing the time required to complete a programming task.
Research shows that students complete tasks faster when using ChatGPT than when using traditional resources, such as textbooks or Stack Overflow~\cite{kosarTomaz, kabirSamia2024}.
However, questions remain about whether faster completion time compromises deeper conceptual learning or fosters students' dependency on the AI tools.
Few studies have measured time efficiency in a counterbalanced classroom experiment with alternating AI access, leaving this dimension underexplored.
This need leads to our second research question.

\begin{tcolorbox}[enhanced,drop shadow]
\textbf{RQ2 asks:  How does ChatGPT usage influence student completion time for programming tasks and post-task assessments?}
\end{tcolorbox}

In computing education, platforms like ChatGPT support novice programmers by offering debugging help, code explanations, and adaptive guidance~\cite{kim2022analysis, raihan2025llmreview, he2024aicompanion}.
Empirical studies show mixed results on the use of ChatGPT.
Some studies report procedural task gains and reduced reliance on traditional resources when structured prompts are provided~\cite{yuankaiICSE_SEET2024, MohammadAlipourAbolnejadianSigcse2024}.
However, other studies note minimal conceptual improvement~\cite{IshikaJoshiSIGCSE2024, shenYiyninAiSIGSCE2024,vassilkaKirovaSIGCSE2024, yuankaiICSE_SEET2024,guner2025ai}.
High-performing students often benefit disproportionately from ChatGPT, suggesting potential ``digital trenches'' in skill development~\cite{lopez2024adoption, qureshi2023chatgpt}.
This prior work motivates our third research question.

\begin{tcolorbox}[enhanced,drop shadow]
\textbf{RQ3: Does access to ChatGPT affect student learning of programming concepts?}
\end{tcolorbox}

Students generally view ChatGPT favorably for improving understanding and engagement~\cite{cipriano2024chatgpt, RogersMichaelSIGCSE2024, kloub2024chatgpt}, but concerns persist about reliability, bias, and ethical use~\cite{stephanieyangICER2024, MengqiLiuSIGCSE2024, ArialHanZhou2024, william2024}.
Research highlights adoption drivers, such as ease of use and feedback quality~\cite{zhang2024futurellm, ALMOGREN2024}, and suggests the use of strategies like oral exams or complex tasks to safeguard academic integrity~\cite{servine2024, mogavi2024chatgpt, kasneci2023chatgpt,kazemitabaar2024codeaid,novak2025, oosterwyk2024hype}.
Yet few studies explicitly connect student perceptions to actual outcomes in controlled lab tasks across multiple programming topics~\cite{Raj2024,Andersen-Kiel2024,oates2025chatgpt,Ouh2025}.
This need leads to our fourth research question.

\begin{tcolorbox}[enhanced,drop shadow]
\textbf{RQ4: What are students’ perceptions about ChatGPT?}
\end{tcolorbox}

Together, these prior findings highlight promise but also gaps, including a lack of counterbalanced designs, limited rubric-based analysis, scarce time-measurement studies, and weak connections between perceptions and actual outcomes.
Our study addresses these gaps by investigating the four research questions listed above that examine performance, code quality, time efficiency, and perceptions in introductory C programming assignments on functions and structures.

\label{related_Work}

\section{Study Design}
This section outlines the study design, including the demographics of the participants, the experimental methodology, the evaluation rubrics, and the data collection procedures.

\subsection{Participants}
We conducted this study in an introductory programming course at a large university in the southeastern United States.
This course, taken primarily by first-semester freshmen, is designed to provide foundational programming skills in the C language.
Twenty-seven students (6 female and 21 male) participated in this study. 
The participants were students enrolled in the CS1 course.
We offered two voluntary lab-style assignments (functions and structures) as part of this study, with minor bonus marks provided as an incentive. 
Students self-enrolled in course sections during routine registration.
Therefore, our assignment of students to groups was a convenience sample..

\subsection{Methodology}
The study employed a quasi-experimental design incorporating pre- and post-surveys, programming assignments, and rubric-based evaluation of student submissions.
We used two assignments (functions, structures) that fit the schedule and provided distinct concepts, balancing ecological validity with feasibility.

To facilitate a controlled comparison, we split the students into two independent groups (Group A and Group B).  
Students in this single introductory CS course self-selected into two sections, which we label Group A and Group B (A: $n=14$, B: $n=13$; the odd class size resulted in one additional student in Group A).
The overall process consisted of four steps:
\begin{enumerate}
    \item Pre-survey administered to both groups to measure baseline attitudes toward AI tools, familiarity with ChatGPT, and confidence in programming skills. 
    \item Students completed two lab-based programming assignments, each focusing on different C concepts (functions and structures).
    \item After each assignment, students completed a post-survey consisting of conceptual assessment questions related to the assignment content and, when applicable, questions about their experience using ChatGPT.
    \item Submitted programs were assessed using a predefined rubric covering correctness, efficiency, readability, and error-handling.
\end{enumerate}
This structured approach enabled a multi-dimensional analysis of ChatGPT’s impact on student learning, procedural skills, and efficiency.

\begin{table*}[!htb]
\centering
\caption{Lab tasks for Assignments 1 and 2, focusing on functions and structures respectively in C programming}
\label{tab:lab_tasks}
{\fontsize{8}{9}\selectfont
\begin{tabular}{l l p{.55\linewidth} l l}
\toprule
& \textbf{Assignment} & \textbf{Lab Task Description} & \textbf{Group A} & \textbf{Group B}\\
\midrule
\multirow{3}{*}{1} & 
\multirow{3}{*}{\textbf{Functions}} & (1) Write a program to check if two strings are anagrams &
\multirow{3}{*}{ChatGPT} & \multirow{3}{*}{No ChatGPT} \\
& & (2) Find the max and min values using a function that returns an array. & & \\
& & (3) Compare two birth years using a function to determine the older person. & & \\
\midrule
\multirow{3}{*}{2} & 
\multirow{3}{*}{\textbf{Structures}} &
(1) Add two distances (inch-feet system) using a structure. & 
\multirow{3}{*}{No ChatGPT} & \multirow{3}{*}{ChatGPT} \\
& & (2) Create a book record using a structure (e.g., author, title, pages, price). & & \\
& & (3) Demonstrate a nested structure example. & & \\
\bottomrule
\end{tabular}
}

\end{table*}

To reduce bias and ensure that all participants experienced both conditions (with and without access to ChatGPT), the study used a counterbalanced design across the two assignments.
Specifically, Group A used ChatGPT for Assignment 1 and did not use it for Assignment 2, while Group B did not use ChatGPT for Assignment 1 and used it for Assignment 2. 
This method of counterbalancing helped to control for any potential effects caused by the order of the assignments, providing more confidence in the results.
Table~\ref{tab:lab_tasks} provides an overview of the assignments.

\subsection{Data Collection}
\label{sec:DataCollection}
We administered the experiment via the Qualtrics platform.
This approach allowed us to present the assignment and gather data all in the same tool.
For each assignment, we gathered the following data.

\subsubsection{Lab Tasks:} 
We collected student code from two programming assignments, each containing three lab tasks designed to assess problem-solving, code quality, and understanding of the target topic.
Assignment 1 focused on functions, and Assignment 2 focused on structures. 
Table~\ref{tab:lab_tasks} lists the specific tasks for each assignment along with group allocations for ChatGPT access.

\subsubsection{Post-Study Surveys}
We used two types of post-study surveys.

\paragraph{Assessment questions} 
This assessment contained questions about the concepts covered in the programming assignment.
The goal was to evaluate the impact of ChatGPT on the students' understanding of the key programming concepts covered in the assignment.

\paragraph{Students' experience and opinion of ChatGPT}
Only the students in the group who used ChatGPT received this survey.
We designed the survey to gather information about students’ experiences with ChatGPT during the assignment, including its perceived helpfulness, frequency of use, and willingness to use it again.

\subsubsection{Time Spent}
We recorded the total time, in seconds, it took each student to complete the programming assignment and to complete and submit the post-assignment survey.
This data provided valuable insights into the efficiency of students' work and enabled a comparative analysis between the group that used ChatGPT and the group that did not use ChatGPT.

\subsection{Evaluation Process of Code Quality Rubrics}
To ensure the accuracy and consistency of the code quality scores assigned to student submissions, two authors independently evaluated the submissions using predefined rubrics. 
These rubrics assessed various aspects of code quality, including correctness, efficiency, code readability, warning analysis, and error-handling shown in Table~\ref{tab_rubrics}.

\begin{table*}[!htb]
\centering
\caption{Rubrics for Assignment 1 (Functions) and Assignment 2 (Structures)}
\label{tab_rubrics}
\small
\setlength{\tabcolsep}{3pt}
\renewcommand{\arraystretch}{0.81}
\begin{tabular}{p{10cm}!{\vrule width 0.8pt}@{\hspace{6pt}}c@{\hspace{6pt}}!{\vrule width 0.8pt}@{\hspace{6pt}}c@{\hspace{6pt}}}
\toprule
\textbf{Category / Criterion} & \textbf{A1} & \textbf{A2} \\
\midrule
\multicolumn{3}{l}{\textbf{Correctness}} \\
Passes all test cases & Q1–Q3 & Q1–Q3 \\
Handles strings, anagram check & Q1 & – \\
Max–min values, all input types & Q2 & – \\
Age comparison incl. leap years & Q3 & – \\
Meaningful error messages & Q3 & – \\
Carry-over in inch–feet addition & – & Q1 \\
Proper book struct + display & – & Q2 \\
Nested struct declaration + display & – & Q3 \\
\midrule
\multicolumn{3}{l}{\textbf{Efficiency}} \\
Time and space efficiency & Q1–Q3 & Q1–Q3 \\
\midrule
\multicolumn{3}{l}{\textbf{Usage}} \\
Relevant struct fields & – & Q2 \\
Nested struct access, meaningful fields & – & Q3 \\
\midrule
\multicolumn{3}{l}{\textbf{Code Quality}} \\
Meaningful names, comments, formatting, structure & Q1–Q3 & Q1–Q3 \\
No unused functions/params/returns & Q2–Q3 & Q1–Q2 \\
\midrule
\multicolumn{3}{l}{\textbf{Error Handling}} \\
Valid and invalid input checks & Q1–Q3 & Q1–Q3 \\
\midrule
\multicolumn{3}{l}{\textbf{Warning Analysis}} \\
No compiler warnings & Q1–Q3 & Q1–Q3 \\
No unused variables & Q1–Q3 & Q1–Q2 \\
Signed/unsigned checks, data loss checks & Q1–Q3 & – \\
No uninitialized variables & Q1–Q3 & Q1–Q3 \\
\bottomrule
\end{tabular}
\end{table*}

\label{studydesign}

\section{Data Analysis and Results}
This section is organized around the four research questions guiding this study, providing a focused examination of ChatGPT's impact on student performance and learning. 
Due to the differences between the two assignments, we analyzed them separately.

\subsection{RQ1: Does using ChatGPT improve the quality of students’ code in lab tasks as measured by our evaluation rubrics?}

\subsubsection{Rubric and Scoring}  
We evaluated each submission using rubrics for both assignments, as shown in Table~\ref{tab_rubrics}.
The rubric items cover five categories: \emph{Correctness, Efficiency, Code Quality, Warning Analysis}, and (for Assignment~2) \emph{Usage of Structures}.
We assigned scores as \textbf{1} (fully met), \textbf{0.5} (partially met), or \textbf{0} (not met).
Two independent raters scored each assignment.
The inter-rater agreement between the two reviewers was $\kappa = 0.87$.
In the cases where we disagreed, we discussed and resolved the disagreements.

The values in Table~\ref{tab:mean_scores_assignments} are the mean scores for each rubric category, calculated across all students in a group for that assignment.
For example, the \textit{Correctness} value for Group A in Assignment 1 is the average correctness score of all Group A students on that assignment.

\begin{table}[!ht]
\centering
\small
\caption{Mean scores by Question and ChatGPT access for Assignment 1 and Assignment 2 with respect to Rubrics}
\label{tab:mean_scores_assignments}
\begin{tabular}{lcc|cc}
\toprule
 & \multicolumn{2}{c|}{\textbf{Assignment 1}} & \multicolumn{2}{c}{\textbf{Assignment 2}}\\
\cmidrule(lr){2-3}\cmidrule(lr){4-5}
\textbf{Category} & \textbf{Group A} & \textbf{Group B} & \textbf{Group A} & \textbf{Group B}\\
 & (GPT) & (No GPT) & (No GPT) & (GPT) \\
\midrule
Correctness        & 2.0 & 1.1 & 2.5 & 1.2 \\
Efficiency / Usage & 2.0 & 1.5 & 1.6 & 1.9 \\
Code Quality       & 2.8 & 1.9 & 2.3 & 1.6 \\
Warning Analysis   & 7.5 & 6.5 & 1.9 & 1.3 \\
Error Handling     & ---   & ---   & 2.0 & 0.05 \\
\bottomrule
\end{tabular}

\end{table}

\subsubsection{Results}

\paragraph{Assignment 1 (Functions).}
The students in Group A (ChatGPT) scored higher than those in Group B (non-ChatGPT).
An independent samples $t$-test confirmed the difference between the groups is significant ($t = 4.46$, $p = 0.018$).

\paragraph{Assignment 2 (Structures).}
After reversing ChatGPT access, Group B (ChatGPT) achieved a higher overall rubric score than Group A (non-ChatGPT).
This difference was also significant ($t = 5.02$, $p < 0.001$).

\paragraph{Cross-assignment interpretation.}
Across both assignments, students with access to ChatGPT recorded significantly higher overall rubric scores, particularly benefiting from improvements in correctness, code readability, and warning analysis.
These findings suggest that AI assistance can help novice programmers improve their code quality, particularly in defensive programming and in applying unfamiliar constructs.

\subsection{RQ2: How does ChatGPT usage influence student completion time for programming tasks and post-task assessments?}
We analyzed completion times for both the lab tasks and the post-survey assessments for each assignment. 

\paragraph{Assignment 1:}  
For the lab tasks, the students in the ChatGPT group finished much faster than the students in the non-ChatGPT group.
This difference was significant ($t = -5.222,\ p = 0.0002$). 
However, for the post-survey assessments, the students in the ChatGPT group had a slightly shorter completion time than those in the non-ChatGPT group. 
The difference was not significant ($t = -0.717,\ p = 0.4828$).

\paragraph{Assignment 2:}  
After switching access, the students in the ChatGPT group again completed the lab tasks faster than those in the the non-ChatGPT group.
The difference was again significant ($t = 4.175,\ p = 0.0011$).  
The students in the ChatGPT group completed the post-survey assessment faster than those in the non-ChatGPT group.
However, unlike in Assignment 1, this difference was significant ($t = 3.996,\ p = 0.0008$).

\medskip
\noindent

\begin{table}[h!]
\centering
\small
\caption{Average completion time (in seconds)}
\label{tab:avgTimePostSurvey}
\begin{tabular}{l|cc|cc}
\toprule
 & \multicolumn{2}{c|}{\textbf{Assignment 1}} & \multicolumn{2}{c}{\textbf{Assignment 2}}\\
\midrule
 & \textbf{Group A} & \textbf{Group B} & \textbf{Group A} & \textbf{Group B}\\
 & (GPT) & (No GPT) & (No GPT) & (GPT) \\
\midrule
\textit{Lab Task}     & 494.92 & 2385.31 & 1934.33 & 868.00 \\
\midrule
\textit{Post-Survey}  & 238.77 & 318.80  & 210.82  & 109.90 \\
\bottomrule
\end{tabular}

\end{table}

\subsection{RQ3: Does access to ChatGPT affect student learning of programming concepts?}
We examined the effect of ChatGPT access on students’ conceptual understanding using the post-assignment survey questions.

\subsubsection{Assignment 1}
Table~\ref{tab:correct_responses_assignmnet1} shows the counts of correct responses to each post-assessment question for Assignment~1.
The Mann--Whitney~U test indicated no statistically significant difference in total correct scores between the ChatGPT and non-ChatGPT groups ($U=153.5$, $p=0.3234$).
This result is driven largely by a substantial advantage on Question~4, ``Uses of functions'' (91.7\% vs.\ 0\%, Odds Ratio $=\infty$, $p<0.0001$).
No other questions showed significant differences: \texttt{fprintf} (OR = NaN, $p=1.0000$), output with functions (OR = 0.173, $p=0.1647$), function arguments (OR = 1.083, $p=1.0000$), and output with recursion (OR = 0.833, $p=1.0000$).

\subsubsection{Assignment 2}
Table~\ref{tab:correct_responses_assignment2} presents the results for Assignment~2.
The Mann--Whitney~U test indicated a statistically significant difference in total correct scores favoring the ChatGPT group ($U=77.0$, $p=0.0457$).
However, none of the per-question Fisher’s exact tests showed significant differences: structure elements storage (OR = 0.300, $p=0.2365$), not possible scenario (OR = 0.455, $p=0.6483$), output with \texttt{for} loop (OR = NaN, $p=1.0000$), structure array initialization (OR = 1.630, $p=0.6776$), and compile successfully with designated initializers (OR = 0.833, $p=1.0000$).

\begin{table}[ht]
\centering
\small
\caption{Assignment 1: Correct Responses of Post-Survey Questions for ChatGPT (Group A) vs Non-ChatGPT (Group B)}
\label{tab:correct_responses_assignmnet1}
\begin{tabular}{p{4cm}cc}
\toprule
\textbf{Question} & \textbf{A (ChatGPT)} & \textbf{B (No ChatGPT)} \\
\midrule
Use of function \texttt{fprintf} in a program & 13 & 14 \\
Output of given C program with functions & 9 & 13 \\
Arguments passed to a function in C language & 1 & 1 \\
Uses of functions & 12 & 0 \\
Output of given C program with recursive function & 5 & 6 \\
\bottomrule
\end{tabular}
\end{table}

\begin{table}[ht]
\centering
\small
\caption{Assignment 2: Correct Responses of Post-Survey Questions for Non-ChatGPT (Group A) vs ChatGPT (Group B)}
\label{tab:correct_responses_assignment2}
\begin{tabular}{p{4cm}cc}
\toprule
\textbf{Question} & \textbf{A (No ChatGPT)} & \textbf{B (ChatGPT)} \\
\midrule
Structure elements storage & 3 & 7 \\
Which of the following is not possible under any scenario & 2 & 4 \\
Output of code with for loop: \texttt{for(i = 0; i < 10; i++); printf("\%d", i);} & 0 & 0 \\
Comment on code with structure array (array initialization) & 4 & 3 \\
Can the following code be compiled successfully (designated initializers in structure)? & 5 & 6 \\
\bottomrule
\end{tabular}
\end{table}

\subsection{RQ4:  What are students’ perceptions about ChatGPT?}
To answer this question, we have data from the pre-survey and the post-survey.

\subsubsection{Pre-Surveys}
Table~\ref{tab:pre_survey_comparison} presents pre-survey results for each assignment.
The percentages indicate the proportion of students in each group that chose the top responses for that category (e.g., ``Very familiar'' or ``Familiar'' for the familiarity row).
The differences for the same group between A1 and A2 reflect that the surveys were administered separately before each assignment, with varying attendance and response rates. 
Thus, values represent only those who responded for that assignment.

Overall, both groups showed high familiarity with ChatGPT, especially in A1. 
Confidence in using ChatGPT and learning the target concepts was moderate to high, slightly higher for functions (A1) than structures (A2). 
Baseline conceptual understanding was generally strong, with occasional lower values in non-ChatGPT groups. These results suggest balanced starting points between groups, minimizing baseline bias.

\begin{table*}[ht]
\centering
\small
\caption{Pre-Survey Comparison Across Groups and Assignments}
\label{tab:pre_survey_comparison}
\begin{tabular}{p{6cm}cccc}
\toprule
\textbf{Category} &
\textbf{A1 ChatGPT (A)} &
\textbf{A1 Non-ChatGPT (B)} &
\textbf{A2 ChatGPT (B)} &
\textbf{A2 Non-ChatGPT (A)} \\
\midrule
Familiarity with ChatGPT (Very familiar / Familiar) & 76.9\% & 60.0\% & 50.0\% & 45.5\% \\
Confidence in ChatGPT (Confident / Very Confident)  & 69.2\% & 80.0\% & 66.7\% & 63.6\% \\
Confidence in learning topic using ChatGPT          & 84.6\% & 86.7\% & 58.3\% & 90.9\% \\
Understanding of concept (Correct conceptual view)   & 92.3\% & 20.0\% & 100\%  & 81.8\% \\
\bottomrule
\end{tabular}
\end{table*}

\subsubsection{Post-Surveys}

\begin{figure}[ht]
  \centering

 \caption{Students' perceptions of ChatGPT by assignment and item.
  Bars show 100\% stacked distributions across response levels (High/Medium/Low).}
  \includegraphics[width=\columnwidth]{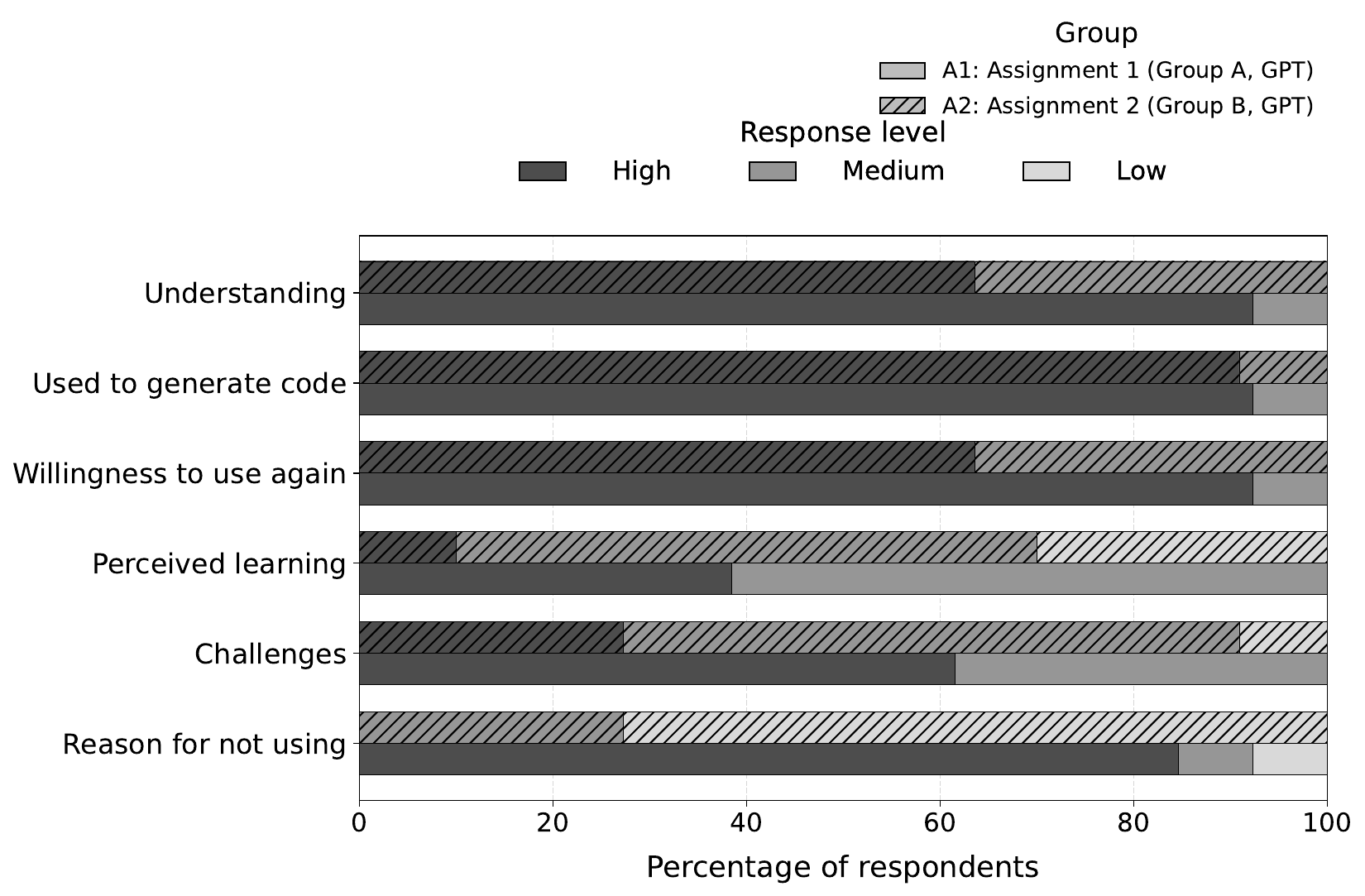}
  \label{fig:chatgpt_perception}
\end{figure}

Figure~\ref{fig:chatgpt_perception} compares the distribution of student responses for each perception category between Assignment 1 (Group A, with ChatGPT) and Assignment 2 (Group B, with ChatGPT).
By specifically crafting questions for these groups, the study aimed to thoroughly assess ChatGPT's perceived usefulness, effectiveness, and overall impact on the learning process.

Across most categories, students in Assignment 1 reported a higher proportion of ``High'' ratings, particularly for understanding the topic, willingness to use ChatGPT again, and perceived learning.
In contrast, Assignment 2 responses showed relatively larger shares of ``Medium'' and ``Low'' ratings in these categories. 
Usage-related items such as writing code and generating code with ChatGPT had consistently high ratings in both assignments, indicating a strong reliance on the tool regardless of assignment order.
The largest divergence between assignments appeared in perceived learning, where Assignment 1 responses skewed more positively than Assignment 2.

\label{data_analysis}

\section{Threats to Validity}
We organize this section around the common types of validity threats.

\subsection{Internal Validity}
Students self-selected into Group A or Group B for the assignments rather than being randomly assigned. 
This self-selection may have introduced selection bias, as we did not control for prior programming experience or technical proficiency, which could have influenced performance differences..
The relatively small sample size increases the likelihood that individual skill variations impacted results.

\subsection{Construct Validity}
Construct validity is threatened by potential misalignment between the study’s evaluation methods and its intended constructs.
While random assignment of students to groups is a fair approach, it is crucial to recognize the impact of differences in prior knowledge, technical skills, and programming aptitude, which can significantly affect the results.
Moreover, enhancing the scoring rubrics used for assignment evaluations will undoubtedly improve their ability to capture critical advancements in conceptual understanding and programming skills, providing a clearer picture of the impact of ChatGPT access. 
Additionally, a careful review of assignment design will allow us to eliminate any unintended advantages for one group over another, particularly if certain tasks are better suited to demonstrating ChatGPT's capabilities.
By proactively addressing these factors, we can achieve a thorough and equitable evaluation of educational outcomes.

\subsection{External Validity}
Our findings are based on students from a single course at a single institution, which limits their generalizability to other contexts, programming languages, or instructional settings. 
The two assignments differed in complexity and topic, which may affect the transferability of results to other tasks. 
Finally, results may not represent long-term effects, as the study measured short-term outcomes within a limited time frame.

\label{threats}

\section{Conclusion}
This counterbalanced, quasi-experimental study examined the influence of ChatGPT on code quality, conceptual understanding, efficiency, and perceptions in a CS1 course.
Across assignments on functions and structures, access to ChatGPT consistently improved rubric-based code quality, particularly in terms of correctness, readability, and warning resolution.
ChatGPT also significantly reduced completion times.
However, effects on conceptual understanding were uneven: in Assignment 1 (Functions), no overall difference emerged except for a substantial gain on a single question, whereas in Assignment 2 (Structures), overall scores favored the ChatGPT group without per-question significance. 
Student perceptions were largely positive, with strong appreciation for its usefulness in code generation and debugging.
Conceptual understanding, measured through post-assignment assessments and compared between groups, showed mixed results: ChatGPT users scored lower for the functions topic but higher for the structures topic, indicating that conceptual benefits may depend on the nature of the topic and task design. 

ChatGPT enhances procedural skills and efficiency; however, structured integration is necessary to support conceptual mastery.
Future work should replicate this study with larger and more diverse populations, extend it to additional programming languages and contexts, and explore longitudinal effects on retention once AI access is removed. 
Systematic variation in task complexity, incorporation of scaffolding strategies such as guided prompts and reflection activities, and fine-grained analysis of AI–student interaction logs could help identify conditions that optimize both procedural and conceptual gains.
Finally, integrating considerations of ethical use and academic integrity will be essential for sustainable pedagogical adoption of generative AI in programming education.

\label{conclusion}
\bibliography{references}
\end{document}